\documentclass[aps,twocolumn]{revtex4}

\begin{document}
\title{Comment on ``Crystal Structure and Pair Potentials: A Molecular-Dynamics
Study''}
\author{Gang Liu}
\address{{\small gang.liu@queensu.ca}\\
{\small High Performance Computing Virtual Laboratory}\\
{\small Queen's University, Kingston, ON K7L 3N6, Canada}}
\date{November 17, 2013}
\maketitle

In 1980, Parrinello~and Rahman proposed a molecular dynamics (PRMD) for
crystal structure under external constant pressure\cite{pr}. Many
simulations of structure phase transitions induced by external forces have
been carried out with it\cite{Mart,Rou,Mor,Tra}, especially after
it was combined with the well-known Car-Parrinello Molecular Dynamics. 
It uses the scaled particle position vectors ${{\bf s}_i}^T$ $%
=\left( \xi _i{\bf ,}\eta _i{\bf ,}\zeta _i\right) $ as generalized coordinates, defined in 
\begin{equation}
{\bf r}_i=\xi _i{\bf a+}\eta _i{\bf b+}\zeta _i{\bf c=Hs}_i\text{, }
\label{c1}
\end{equation}
where ${\bf r}_i$ is the $i$-th particle position vector, $\ {\bf a}${\bf , }$%
{\bf b}$ and{\bf \ }${\bf c}$ are the cell edge vectors, the matrix ${\bf H}%
=\left\{ {\bf a,b,c}\right\} $, and $T$ is the transpose.
In physics, it should make no difference by using either ${\bf s}_i$ or ${\bf r}_i$. 
PRMD further uses 
\begin{equation}
K^{\prime }=\text{ }\sum_i\frac 12m_i \dot {{\bf s}_i}^T{\bf H}^T{\bf H%
} \dot {{\bf s}_i}  \label{c2}
\end{equation}
as the particles' kinetic energy, while the true kinetic energy for
particles is 
\begin{eqnarray}
K &=&\text{ }\sum_i\frac 12m_i\dot {{\bf r}_i}^2  \nonumber
 \\
&=&\text{ }\sum_i\frac 12m_i\left( \dot {{\bf s}_i}^T{\bf H}^T+{\bf s}%
_i^T\dot {\bf H}^T\right) \left( {\bf H}\dot {{\bf s}_i}+%
\dot {\bf H}{\bf s}_i\right) ,  \label{c3}
\end{eqnarray}
where $m_i$ is the $i$-th particle's mass. The PRMD starts
from the Lagrangian $L$ introduced as 
\begin{equation}
L=K^{\prime }-\Phi +L_{{\bf H}},  \label{c6}
\end{equation}
where $\Phi $ is the potential energy among particles and $L_{{\bf H}}$ is
the Lagrangian associated with $\dot {\bf H}$, ${\bf H}$, and
external pressure and $L_{{\bf H}}$ has no contribution to the particles'
dynamical equations. The purpose of this comment is to study whether the chosen
kinetic energy is reasonable by comparing the generated dynamical equations for
particles with the Newton's Second Law 
\begin{equation}
m_i\ddot {{\bf r}_i}={\bf F}_i\text{,}  \label{c5}
\end{equation}
where ${\bf F}_i$ is the net force acting on particle $i$. Now let us 
do the following three cases in sequence.

Case 1, from the Lagrangian dynamical equation, we can easily reproduce 
Eq.~(\ref{c5}) by using ${\bf r}_i$ as ``generalized coordinates'' and using the
complete kinetic energy Eq.~(\ref{c3}) to replace $K^{\prime }$ in Eq.~(\ref
{c6}), with 
\begin{equation}
{\bf F}_i=-\frac {\partial \Phi} {\partial {\bf r}_i} \text{.}  \label{c7}
\end{equation}

Case 2, let us use ${\bf s}_i$ as generalized coordinates and still use the
complete kinetic energy of Eq.~(\ref{c3}). Combining Eq.~(\ref{c1}) and (\ref{c7}%
), we obtain $-\frac \partial {\partial {\bf s}_i}\Phi ={\bf H}^T{\bf F}_i$%
. By also using of $\frac \partial {\partial {\bf s}_i}K=m_i\dot {\bf %
H}^T\left( {\bf H}\dot {{\bf s}_i}+\dot {\bf H}{\bf s}%
_i\right) $ and $\frac \partial {\partial \dot {{\bf s}_i}}K=m_i{\bf H%
}^T\left( {\bf H}\dot {{\bf s}_i}+\dot{\bf H}{\bf s}_i\right) 
$, the Lagrangian dynamical equation, $\frac d{dt}\frac \partial {\partial 
\dot {{\bf s}_i}}L=\frac \partial {\partial {\bf s}_i}L$, becomes  
\begin{equation}
m_i{\bf H}^T\left( {\bf H}\ddot {{\bf s}_i}+2\dot {\bf H}%
\dot {{\bf s}_i}+\ddot{\bf H}{\bf s}_i\right) ={\bf H}^T{\bf F%
}_i,  \label{c13}
\end{equation}
which is equivalent to Eq.~(\ref{c5}).

Case 3, as in the PRMD, let us use ${\bf s}_i$ as generalized coordinates
and use the incomplete kinetic energy of Eq.~(\ref{c2}). Similar to the
above case, with $\frac \partial {\partial \dot {{\bf s}_i}}K^{\prime
}=m_i{\bf H}^T{\bf H}\dot {{\bf s}_i}$, the particle's dynamical
equation becomes 
\begin{equation}
m_i\left( {\bf H}^T{\bf H}\ddot {{\bf s}_i}+{\bf H}^T\dot {\bf %
H}\dot {{\bf s}_i}+\dot{\bf H}^T{\bf H}\dot {{\bf s}_i}%
\right) ={\bf H}^T{\bf F}_i,  \label{c15}
\end{equation}
which is the same as the Eq.~(2) of the PRMD's original paper\cite{pr}, but
different from Eq.~(\ref{c5}). The discrepancy can be found in $%
\frac d{dt}\frac \partial {\partial \dot {{\bf s}_i}}\left( K-K^{\prime }\right) 
$ and $\frac \partial {\partial {\bf s}_i}\left( K-K^{\prime }\right) $. So
the incomplete kinetic energy of Eq.~(\ref{c2}) is not reasonable.

If the complete kinetic energy of Eq.~(\ref{c3}) is used, the first term on
the right-hand side of Eq.~(4) of the PRMD's original paper\cite{pr} will disappear
in physics. Anyhow, the
PRMD is correct for crystal structure when all accelerations and
velocities are zero.


\begin{references}
\bibitem{pr}  M. Parrinello,~and A. Rahman, Phys. Rev. Lett. {\bf 45,} 1196
(1980).

\bibitem{Mart}  R. Martonak, {\it et al.} , Phys. Rev. Lett. {\bf 84,} 682 (2000).

\bibitem{Rou}  R. Rousseau, {\it et al.} , Phys. Rev. Lett. {\bf 85,} 1254 (2000).

\bibitem{Mor}  T. Morishita, Phys. Rev. Lett. {\bf 87,} 105701 (2001).

\bibitem{Tra}  A. Trave, {\it et al.}, Phys. Rev. Lett. {\bf 89,} 245504 (2002).

\end{references}
\end{document}